\documentclass{article}
\usepackage{frascatiphys_C}
\begin{document}
\title{ 
RECENT BES RESULTS ON $\psi^{\prime}$ DECAY }
\author{
 XiaoHu Mo \thanks{\hspace{0.2cm}  On behalf of BES collaboration }\\
\em Institution of High Energy Physics, CAS, Beijing 100039, China }
\maketitle
\baselineskip=11.6pt
\begin{abstract}
With 14 M $\psi^{\prime}$ events, many two-body decay channels are studied, 
which include VP, VT and PP channels. Based on systematical measurements
for charmonium decay, 12\% rule is tested, the phase between strong and
EM amplitudes is studied. In addition, hadronic and radiative transition of
charmonia are measured to improve experimental accuracy and test theoretical 
calculations.
\end{abstract}
\baselineskip=14pt
\section{Introduction}
Charmonium decay continues to present itself as a challenge to our 
understanding of the strong interaction. Up to 2004, BES collaboration has 
collected 14 Million (M) $\psi^{\prime}$ events (luminosity is 19.72 pb$^{-1}$), 58 M $J/\psi$ events, 27 pb$^{-1}$ $\psi^{\prime \prime}$ data and 6.4 
pb$^{-1}$ data taken at 3.65 GeV for continuum study. With all these samples, 
studies have made systematically for charmonium decay. Herein the results of 
$\psi^{\prime}$ decay is the main content of this report, which contains 
the following topics: decays of $\psi^{\prime}$
to Vector Pseudoscalar (VP), Vector Tensor (VT), Pseudoscalar Pseudoscalar (PP) channels, and hadronic and radiative transition of $\psi^{\prime}$.

As it is known, both $J/\psi$ and $\psi^{\prime}$ decays are expected to be 
dominated by annihilation into three gluons, with widths that are proportional 
to the square of the $c\bar{c}$ wave function at the origin~\cite{appelquist}. 
This yields the pQCD expectation (so-called ``12 \% '' rule) that
\begin{equation}
Q_h =\frac{{\cal B}_{\psi^{\prime} \rightarrow X_h}}
          {{\cal B}_{J/\psi \rightarrow X_h}}
=\frac{{\cal B}_{\psi^{\prime} \rightarrow e^+e^-}}
      {{\cal B}_{J/\psi \rightarrow e^+e^-}}
= (12.3 \pm 0.7) \%~~.
\label{pqcdrule}
\end{equation}
The observation of deviation from 12 \% rule will 
provide some new clues concerning the dynamics of charmonium decay. Another 
study relevant to charmonium decay is the relative phase $\phi$ between 
strong and electromagnetic (EM) amplitudes. At $J/\psi$ region, the nature of
$\phi$ has been studied in many two-body decay modes: 
$1^-0^-$~\cite{dm2exp,mk3exp}, $0^-0^-$~\cite{a00,lopez,kopke}, 
$1^-1^-$~\cite{kopke} and $N\overline{N}$~\cite{ann}; 
while at $\psi^{\prime}$ region, only two modes $0^-0^-$~\cite{ppphase} and
$1^-0^-$~\cite{vpphase} have been discussed phenomenologically, more researches are needed.

Here it is necessary to stress a point. In $e^+e^-$ experiment, the production
of $\psi^{\prime}$ is accompanied by one photon continuum process
\begin{equation}
e^+e^- \rightarrow \gamma^* \rightarrow \mbox{hadrons}~,
\end{equation}
in which $e^+e^-$ pair annihilates into a virtual photon without going through
the intermediate resonance state. Taking the contribution from this
process and its interference effect into consideration, it could
determine not only the magnitude but also the sign of $\phi$. Furthermore,
the continuum contribution and its interference effect will exert obvious
influence on the branching ratio measurement, which should be treated
carefully in corresponding analyses.

\section{Study of $\psi^{\prime}$ two-body decay}
\subsection{VP channel}
\begin{figure}[htb]
\vspace{5.0cm}
\begin{minipage}{5cm}
\includegraphics{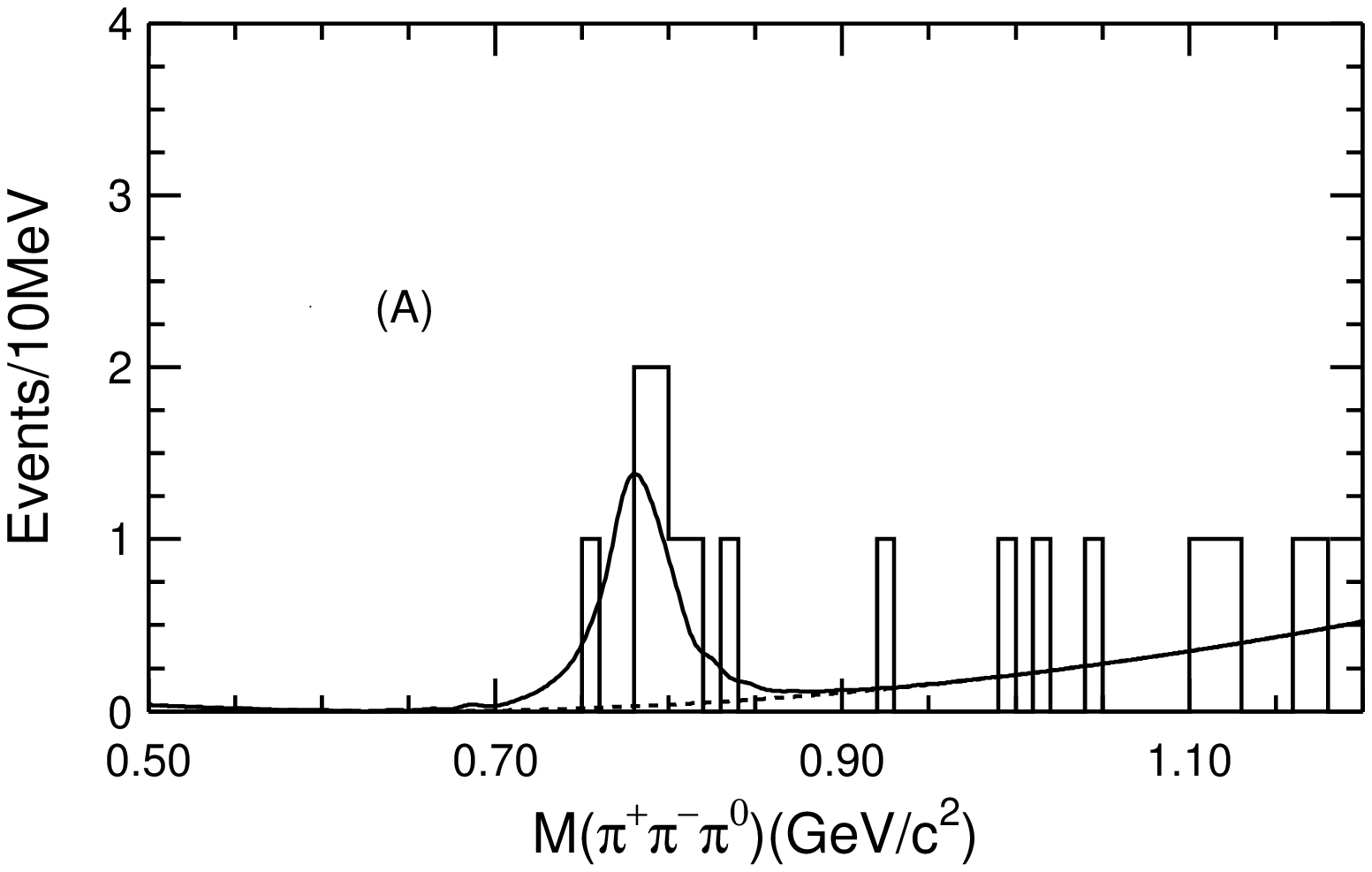}
\mbox{} \hspace{1.0cm} (a) At $E_{cm}=3.65$ GeV
\end{minipage}
\begin{minipage}{7cm}
\includegraphics{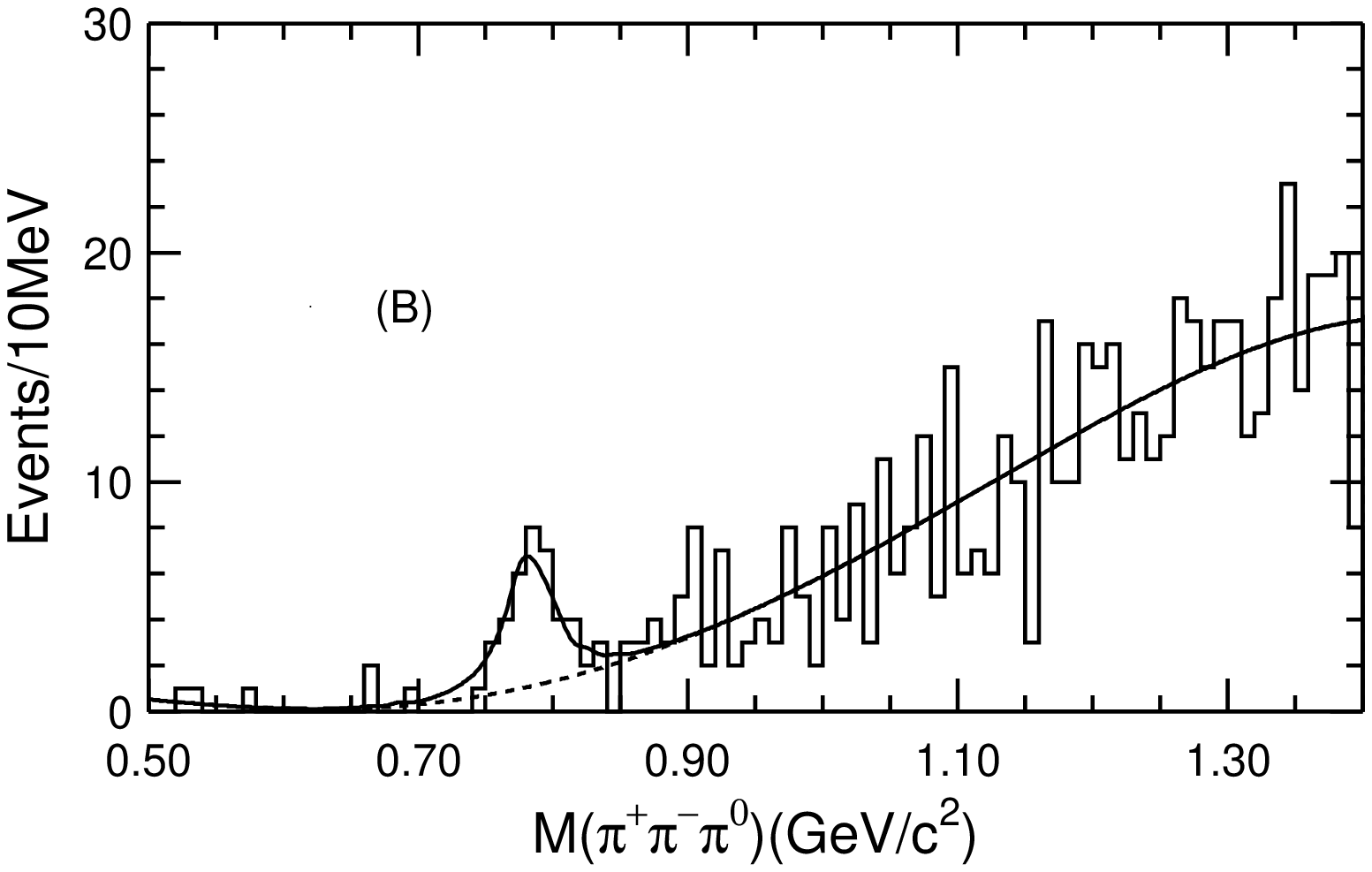}
\mbox{} \hspace{2.0cm} (b) At $E_{cm}=3.686$ GeV
\end{minipage}
\caption{\it The invariant mass distribution of $\omega$ at (a) continuum
and (b) resonance. The dashed line indicates the background while
the solid line the synthetic fitting result. \label{psipop} }
\end{figure}

As forementioned the continuum contribution need to be treated carefully, 
the data at both resonance and continuum are analyzed. Fig. \ref{psipop} shows 
the invariant mass distribution of $\omega$, from which the numbers of events 
are fitted to be $7.4 \pm 2.8$ at $E_{cm}=3.65$ GeV and $31.3\pm 7.4$
at $E_{cm}=3.686$ GeV, respectively. The rough estimation based on the 
present results shows the continuum contribution is around 70\%, which is 
consistent with 60\%, the phenomenological calculation~\cite{fofac}.
For $K^* \overline{K}$ channel, $K \pi K_S $
($K_S \rightarrow \pi^+ \pi^-$) final state is studied. From the invariant mass
distributions of $K\pi$ and $K_S \pi$ at $\psi^{\prime}$ peak (continuum), the 
numbers of events are fitted to be $65.6 \pm 9.0$ ($2.5 \pm 1.9$) and 
$9.6 \pm 4.2$ ($~0~$) for $K^* \overline{K}^0+c.c.$ and 
$K^{*+} \overline{K}^- +c.c.$ respectively. 
With the luminosities, it is easy to transform the observed numbers of events 
into the corresponding cross sections. If the parameterization forms
in reference\cite{kopke} are adopted, and observed cross sections are used as
inputs, the phase between strong and EM amplitudes can be fitted out, at the 
same time, obtained are the branching ratios, which are $12.7\times 10^{-5}$ 
and $3.1\times 10^{-5}$ for $K^* \overline{K}^0+c.c.$ and 
$K^{*+} \overline{K}^- +c.c.$, respectively. Comparing with the results listed
in Table~\ref{psipres}, from which the continuum contribution has not been 
subtracted, the largest difference is around 18\%. 

\subsection{VT channel}
The measured results for VT channel~\cite{besvt} 
are listed in Table~\ref{psipres}, from which we notice 
the Q-value for all VT channel are suppressed by a factor of 3 to 5
compared with the 12 \% rule.

\subsection{PP channel}
For PP channel, the parameterization forms~\cite{haber}
\begin{equation}
\begin{array}{rcr}
\pi^+ \pi^-         :&              &   E~,     \\
 K^+ K^-            :& \sqrt{3/2}~M &+~E~,     \\
  K^0_S K^0_L       :& \sqrt{3/2}~M &,~~~~~ 
\end{array}
\end{equation}
are adopted to determine the phase $\phi$. So far as $e^+e^-$ experiment is 
concerned, $E$ must be replaced by $E+E_C$, where $E_C$ denotes the continuum 
contribution. With measurements before~\cite{ppphase} and the recently measured 
branching ratio for $\psi^{\prime} \rightarrow K^0_S K^0_L$, we can fit out 
$\phi$ to be $(-82\pm 29)^{\circ}$ or $(+121\pm 27)^{\circ}$. The detailed 
analyses of $K^0_S K^0_L$ in $J/\psi$ and $\psi^{\prime}$ decay can be found in 
references~\cite{jpsikskl} and ~\cite{psipkskl}, the final results 
are summarized in Table~\ref{psipres}. 
\begin{table}[htb]
\centering
\caption{\it The results of $\psi^{\prime}$ two-body decay.}
\vskip 0.1 in
\begin{tabular}{cccc} \hline \hline
 VP channel &  ${\cal B}_{\psi^{\prime}}~(10^{-5})$ 
            &  ${\cal B}_{J/\psi}~(10^{-4})$ & $Q_h$ \\ 
            &  (from BES)       
	    &  (from PDG2002)                &       \\ \hline
$K^* \overline{K}^0 + c.c.$
            & $15.0 \pm 2.1 \pm 1.7$    
            & $ 42 \pm 4           $       & $3.6 \pm 0.7$ \\
$K^{*+} \overline{K}^- + c.c.$
            & $ 2.9 \pm 1.3 \pm 0.4$     
            & $ 50 \pm 4           $       & $0.58 \pm 0.29$ \\
$\omega \pi^0$  
            & $< 3.27$
	    & $4.2 \pm 0.6         $       & $< 7.8$ \\  \hline \hline
 VT channel &  ${\cal B}_{\psi^{\prime}}~(10^{-4})$ 
            &  ${\cal B}_{J/\psi}~(10^{-3})$ & $Q_h$ \\ 
            &  (from BES)       
	    &  (from PDG2002)                &       \\ \hline
$\omega f_2$& $2.05 \pm 0.41 \pm 0.38$  
            & $4.3  \pm 0.6          $       & $4.8 \pm 1.5$ \\
$\rho a_2$  & $2.55 \pm 0.73 \pm 0.47$    
            & $10.9 \pm 2.2          $       & $2.3 \pm 1.1$ \\
$K^* \overline{K^*_2} + c.c.$
            & $1.86 \pm 0.32 \pm 0.43$    
            & $6.7  \pm 2.6          $       & $2.8 \pm 1.3$ \\
$\phi f_2^{\prime}$  
            & $0.44 \pm 0.12 \pm 0.11$
	    & $1.23 \pm 0.21         $ 
	                                     & $3.6 \pm 1.5$ \\  \hline \hline
 PP channel & ${\cal B}_{\psi^{\prime}}~(10^{-5})$ 
            & ${\cal B}_{J/\psi}~(10^{-4})$   & $Q_h$ \\ 
            &  (from BES)         
	    &  (from BES)                    &       \\ \hline
$K^0_SK^0_L$& $5.24 \pm 0.47 \pm 0.48$
            & $1.82 \pm 0.04 \pm 0.13$       & $28.8 \pm 3.7$ \\ \hline  \hline
\end{tabular}
\label{psipres}
\end{table}

\section{12\% rule and mixing model}
The Q-values for three kinds of two-body decay, VP, VT and PP, are listed in 
Table~\ref{psipres}. It shows clearly the Q-value is enhanced for some 
channels while suppressed for others. In fact, many theoretical efforts are 
made to settle the problems~\cite{rhopitheo}, however, none explains all 
the existing experimental data naturally. Here we only mention one point: 
some recent phenomenological studies indicate that S- and D-wave mixing model 
is a natural and calculable model. It probably give a unified explanation for 
all 12\% rule deviated decays. Using this model, according to the measurement
results at $J/\psi$ and $\psi^{\prime}$, the corresponding decay at 
$\psi^{\prime \prime}$ can be predicted. So the measurement at 
$\psi^{\prime \prime}$ can be used to test the mixing model.
One example is given in reference~\cite{psippkskl}, according to which 
the branching ratio of $\psi^{\prime \prime} \rightarrow  K^0_S K^0_L$ 
is estimated to be within a range from $(0.12 \pm 0.07) \times 10^{-5}$
to $(3.8 \pm 1.1) \times 10^{-5}$.
With the data at $\psi^{\prime \prime}$, BES has detected an upper limit, 
which does not contradict with the current prediction.

\section{$\psi^{\prime}$ hadronic and radiative transition}
Motivation for such study is to improve experimental accuracy and
test theoretical calculations. Inclusive and exclusive methods are adopted
to analyze the following channels extensively:

\begin{equation}
\begin{array}{cc}
X J/\psi(J/\psi \rightarrow \mu^+ \mu^-) \mbox{final state} &
\gamma \gamma J/\psi(J/\psi \rightarrow \ell^+\ell^-)
 \mbox{final state} \\
   Anything J/\psi   &  \pi^0 J/\psi     \\
\pi^0 \pi^0 J/\psi   &  \eta J/\psi      \\
       \eta J/\psi   &  \gamma \chi_{c1}, 
                               \chi_{c1} \rightarrow \gamma J/\psi \\
 \gamma \chi_{c1}, 
        \chi_{c1} \rightarrow \gamma J/\psi 
	             & \gamma \chi_{c2}, 
		              \chi_{c2} \rightarrow \gamma J/\psi \\
 \gamma \chi_{c2}, 
        \chi_{c2} \rightarrow \gamma J/\psi
	             &               
\end{array}
\end{equation}
For $X J/\psi$ final states, $\mu$-pair
is used to identify $J/\psi$ particle, 
the invariant mass distributions of X with and without extra charged-track 
cases are fitted simultaneously with component shapes determined from Monte
Carlo simulation~\cite{xjpsi}; for $\gamma \gamma J/\psi$ final states, 
lepton-pair is used to identify $J/\psi$ particle, the various exclusive 
channels are fitted separately~\cite{ggjpsi}.
Based on BES results, some theoretical calculations are tested. 
Comparisons show the calculation based on PCAC are smaller than BES 
measurement, while the Multipole expansion evaluations are consistent with 
BES present values~\cite{ggjpsi}.
\section{Acknowledgments}
Thanks my colleagues of BES collaboration who provide me so many good results
which are reported here. 
\end{document}